\documentclass[lettersize,journal]{IEEEtran}
\usepackage{amsmath,amsfonts}
\usepackage{algorithmic}
\usepackage{algorithm}
\usepackage{array}
\usepackage{graphicx}
\usepackage[caption=false,font=normalsize,labelfont=sf,textfont=sf]{subfig}
\usepackage{textcomp}
\usepackage{stfloats}
\usepackage{url}
\usepackage{verbatim}
\usepackage{cite}
\usepackage{multirow}
\usepackage{hyperref}
\begin{document}

\title{Efficient Sparse Coding with the Adaptive Locally Competitive Algorithm for Speech Classification}

\author{Soufiyan Bahadi, Eric Plourde, and Jean Rouat}


\maketitle
\begin{abstract}
Researchers are exploring novel computational paradigms such as sparse coding and neuromorphic computing to bridge the efficiency gap between the human brain and conventional computers in complex tasks. A key area of focus is neuromorphic audio processing. While the Locally Competitive Algorithm has emerged as a promising solution for sparse coding, offering potential for real-time and low-power processing on neuromorphic hardware, its applications in neuromorphic speech classification have not been thoroughly studied. The Adaptive Locally Competitive Algorithm builds upon the Locally Competitive Algorithm by dynamically adjusting the modulation parameters of the filter bank to fine-tune the filters' sensitivity. This adaptability enhances lateral inhibition, improving reconstruction quality, sparsity, and convergence time, which is crucial for real-time applications. This paper demonstrates the potential of the Locally Competitive Algorithm and its adaptive variant as robust feature extractors for neuromorphic speech classification. Results show that the Locally Competitive Algorithm achieves better speech classification accuracy at the expense of higher power consumption compared to the LAUSCHER cochlea model used for benchmarking. On the other hand, the Adaptive Locally Competitive Algorithm mitigates this power consumption issue without compromising the accuracy. The dynamic power consumption is reduced to a range of $4$ to $13$ milliwatts on neuromorphic hardware, three orders of magnitude less than setups using Graphics Processing Units. These findings position the Adaptive Locally Competitive Algorithm as a compelling solution for efficient speech classification systems, promising substantial advancements in balancing speech classification accuracy and power efficiency.
\end{abstract}


\section{Introduction}
\IEEEPARstart{T}{he} human brain, despite its billions of neurons, exhibits remarkable computational efficiency, surpassing conventional computers in tasks like pattern recognition and adaptation. Understanding this efficiency gap motivates researchers to explore novel computational paradigms inspired by the brain's structure and function. Two promising avenues in this pursuit are sparse coding and neuromorphic computing.

Sparse coding, a concept deeply rooted in neuroscience, refers to a coding strategy wherein representations are characterized by a sparse set of active units, with the majority of units remaining inactive in response to stimuli \cite{sparse1, sparse2}. This coding scheme has been observed across various sensory modalities and brain regions, suggesting its fundamental role in information processing and efficient neural communication \cite{sparsecomm1, sparsecomm2}. In recent years, sparse coding has emerged as a prominent framework in computational neuroscience and machine learning, offering insights into the organization of neural circuits and inspiring novel algorithms for data representation and analysis \cite{repr1, repr2}.

Concurrently, neuromorphic computing has gained momentum as a transformative approach to computing, drawing inspiration from the architecture and dynamics of biological nervous systems \cite{neuromorphic1, neuromorphic2}. Unlike conventional computing paradigms, which rely on the von Neumann architecture and sequential processing, neuromorphic systems emulate the massively asynchronous, parallel, and distributed nature of the brain, enabling efficient and low-power information processing \cite{lowp1, lowp2}. By leveraging principles of sparsity and neural computation, neuromorphic architectures hold the potential to unlock new frontiers in artificial intelligence, enabling autonomous learning, adaptive behaviour, and energy-efficient computing \cite{eecomp1, eecomp2}.

Sparse coding and neuromorphic computing are not isolated endeavours. In fact, sparse coding algorithms can be implemented on neuromorphic hardware, potentially unlocking significant efficiency gains. Since sparse coding involves fewer active components compared to dense representations, they typically require less energy for computation and communication. This aligns well with the goal of neuromorphic computing to develop energy-efficient and brain-like AI systems.

One prominent example is the Locally Competitive Algorithm (LCA) \cite{lca}, a sparse coding implementation that is highly compatible with neuromorphic hardware. LCA maps sparse representation features to neurons that compete to sparsely reconstruct the input, resulting in neuromorphic implementations such as \cite{slcaloihi1, slcaloihi2}. LCA was initially developed for image processing and has been successfully applied to tasks such as denoising \cite{denoise}, up-sampling \cite{upscale}, compression \cite{compress}, and image classification \cite{lca, im1, im2} among others. Although LCA has shown promise in audio signal processing, particularly with the introduction of Gammatone and compressive Gammachirp filter banks \cite{plca, clca}, its applications in audio processing remain limited. Specifically, in the context of neuromorphic speech classification, LCA has not yet been explored as a feature extractor for spiking neural networks (SNNs).

In previous work \cite{alca}, we leveraged LCA's neural architecture to develop the Adaptive LCA (ALCA), a framework that dynamically adjusts the modulation parameters of Gammachirps using the backpropagation algorithm. This adaptation mechanism fine-tunes the sensitivity of filters around their central frequencies based on the acoustical environment. While the backpropagation algorithm deviates from biological processes, its implementation in ALCA parallels certain biological mechanisms: the Gammachirp filter bank approximates the frequency analysis of the basilar membrane, and the application of backpropagation to influence the filter’s sensitivity around the central frequency mirrors the efferent modulation observed in hair cell function. ALCA improves reconstruction quality and sparsity of extracted features while reducing convergence time, an essential aspect for real-time applications. However, the impact of ALCA’s sparsity improvements on neuromorphic speech classification remains unexplored.

In this article, we extend the applicability of LCA and evaluate it in the context of neuromorphic speech classification. Two tasks are considered in this study: the Heidelberg Digits and Speech Commands classification tasks. These datasets have been encoded into spikes using a software silicon cochlea model named LAUSCHER \cite{shd} and publicly shared for SNNs benchmarking purposes. Additionally, we evaluate the impact of the adaptation introduced by ALCA on the performance of these tasks. Specifically, we conduct a comprehensive comparison between ALCA, LCA, and LAUSCHER based on representation sparsity, the classification accuracy of the processing SNN, and spiking activity. We also assess ALCA/LCA’s performance on temporal tasks to highlight the importance of temporal information in the representation. Through this detailed comparison, we aim to clarify the strengths and weaknesses of each approach in two different speech classification contexts. Furthermore, we explore the implications of ALCA's improved reconstruction quality and sparsity on speech classification accuracy and neural network activity, validating its efficacy in extracting useful features from audio signals. Finally, we compare the power consumption of neural networks using the three representations (ALCA, LCA, and LAUSCHER cochlea model) on both GPU and Intel’s research Loihi 2 chip \cite{loihi2}.

In summary, the key contributions of this article are:

\begin{itemize}
    \item Extending the Applicability of ALCA / LCA: we explore the ability of ALCA and LCA to extract features directly from audio signals for use in neuromorphic speech classification.
    \item Comparison between ALCA, LCA, and LAUSCHER cochlea model: we conduct a comprehensive comparison between ALCA, LCA and LAUSCHER based on representation sparsity, classification accuracy of SNNs, and spiking activity.
    \item Evaluation of the impact of ALCA’s adaptation on the performance and power consumption in neuromorphic speech classification.
\end{itemize}

Our results demonstrate that ALCA emerges as the most efficient sparse representation method compared to the considered representations. Notably, it not only achieves the most accurate speech classification but is also the most sparse representation among the three evaluated methods. Moreover, our findings reveal that ALCA requires minimal power consumption for speech classification tasks, particularly evident when implemented on Intel's Loihi 2 chip\cite{loihi2}. These insights underscore the significant advantages of the ALCA representation in both accuracy and power efficiency, positioning it as a promising solution for neuromorphic speech processing applications.

\section{METHODS}
\label{sec:method}

\subsection{Speech Representations}
\label{ssec:sparse}
In this section, we will describe briefly the three types of speech representations that we evaluate. Namely, LAUSCHER, LCA and ALCA.

\subsubsection{LAUSCHER Cochlea Model Representation}
\label{sssec:shd/ssc}
This representation serves as a reference representation to which all other representations in this study are compared. It is a spike-based representation generated by a software silicon cochlea model that has been publicly shared in \cite{shd}. 
The software models the inner ear and parts of the ascending auditory pathway. This biologically inspired model performs similar steps to produce spike-based representation as in \cite{meddis}. First, a hydrodynamic basilar membrane model causes spatial frequency dispersion. Which is comparable to a frequency analysis by computing a $700$-channel time-frequency representation with a Mel-spaced filter bank.
Second, these separated frequencies are converted to instantaneous firing rates through a transmitter pool-based hair cell model inspired by biology, which adds refractory effects and a layer of bushy cells that increase phase locking. All model parameters were chosen to mimic biological findings.

\subsubsection{Locally Competitive Algorithm (LCA) for Audio}
\label{sssec:lcaaudio}
The second representation is the LCA with Gammachirp dictionary as described in \cite{alca, plca}. The goal of LCA is to represent input signals as a linear combination of a family of vectors (atoms) $\boldsymbol{D} = [\boldsymbol{\phi}_1, ..., \boldsymbol{\phi}_N]$, called ``dictionary'', where most coefficients $\boldsymbol{a} = [a_1, ..., a_N]^T$ are zero:
\begin{equation}
    \label{eq:recons}
    \displaystyle
    \boldsymbol{\hat{s}} = \sum_{m=1}^{N} a_{m} \boldsymbol{\phi}_m = \boldsymbol{D}\boldsymbol{a},
\end{equation}
where $\boldsymbol{\hat{s}}$ is the approximation of the input signal $\boldsymbol{s}$ and $N$ is the number of atoms. To derive the coefficients $\boldsymbol{a}$, a recurrent neural network incorporating lateral inhibition is defined with an objective function to be minimized. This function is referred to as an energy function $E$ defined as a combination of the Mean Squared Error (MSE) between $\boldsymbol{s}$ and $\boldsymbol{\hat{s}}$, a sparsity cost penalty $S$ evaluated from the activation of neurons that corresponds to the coefficients $\boldsymbol{a}$ in (\ref{eq:recons}), and a Lagrange multiplier $\lambda$:
\begin{equation}
    \label{eq:energy}
    \displaystyle
    E = \frac{1}{2}||\boldsymbol{\hat{s}}-\boldsymbol{s}||^2 + \lambda S(\boldsymbol{a}).
\end{equation}

The key characteristic of LCA that allows the minimization of $E$, as will be discussed further below, is its neural dynamics. These dynamics are governed by the vectorized ordinary differential equation:
\begin{equation}
    \label{eq:ODE}
    \displaystyle
    \tau \frac{d\boldsymbol{v}}{dt} = \boldsymbol{p} - \boldsymbol{v} - (\boldsymbol{D}^T\boldsymbol{D}-\boldsymbol{I})\boldsymbol{a},
\end{equation}
where $\tau$ is the time constant of each neuron, $\boldsymbol{p}$ is the input signal projection on the dictionary, i.e., $\boldsymbol{p} = \boldsymbol{D}^T \boldsymbol{s}$, $\boldsymbol{v}$ is the membrane potential vector, and $\boldsymbol{I}$ is the identity matrix. Basically, the evolution of $\boldsymbol{v}$ over time depends on the input intensity $\boldsymbol{p}$ and on $-\boldsymbol{v}$ which makes these neurons behave like leaky integrators. Membrane potentials exceeding the threshold $\lambda$ –––the same as the Lagrange multiplier in (\ref{eq:energy})––– produce activations corresponding to the coefficients $\boldsymbol{a}$. Each activated neuron inhibits all others through horizontal connections, where $\boldsymbol{D}^T\boldsymbol{D}-\boldsymbol{I}$ are the weights of these connections. For each neuron $m$, the activation $a_m$ is a non-linearity $T_\lambda$ that can be sigmoidal or –––as we used in this work––– the hard thresholding function applied to the potential $v_m$ of neuron $m$:
\begin{equation}
    \label{eq:thresh}
    \displaystyle
    a_m = T_\lambda(v_m) =
    \begin{cases}
      0, & \text{if $|v_m| < \lambda$}\\
      v_m, & \text{otherwise}
    \end{cases}.
\end{equation}
It has been shown in \cite{lca} that by imposing the following relation between activations $\boldsymbol{a}$, potentials $\boldsymbol{v}$, and the sparsity cost $S$,
\begin{equation}
    \label{eq:cost}
    \displaystyle
    \lambda\frac{\partial S(\boldsymbol{a})}{\partial a_m} = v_m - a_m
\end{equation}
the evolution over time of $\boldsymbol{v}$ becomes $\frac{d\boldsymbol{v}}{dt} \propto -\frac{\partial E}{\partial \boldsymbol{a}}$. The energy function (\ref{eq:energy}) is therefore minimized.
Using equations (\ref{eq:thresh}) and (\ref{eq:cost}), one can show that $S$ is proportional to the $L_0$ norm such that $S(\boldsymbol{a}) = \frac{\lambda}{2} L_0(\boldsymbol{a})$. For more details, refer to \cite{lca}.

Minimizing (\ref{eq:energy}) with respect to the coefficients $\boldsymbol{a}$ is thus equivalent to solving (\ref{eq:ODE}). The solution of (\ref{eq:ODE}) in discrete time for time step $n$ and step size $\Delta t$ is:
\begin{align}
    \label{eq:euler}
    \displaystyle
    \begin{aligned}
        \boldsymbol{v}[n] = & \frac{\Delta t}{\tau} (\boldsymbol{D}^T \boldsymbol{s} - (\boldsymbol{D}^T\boldsymbol{D}-\boldsymbol{I}) \boldsymbol{a}[n-1])\\ & +(1-\frac{\Delta t}{\tau})\boldsymbol{v}[n-1]
    \end{aligned}.
\end{align}
In the following experiments we used $\frac{\Delta t}{\tau} = 0.01$ for $64$ iterations as in \cite{alca}.

The dictionary $\boldsymbol{D}$ is composed of Gammachirp filters impulse responses \cite{Gammachirp} which are formed by the combination of a monotonically frequency modulated carrier—a chirp—and an envelope that is taken as a Gamma distribution function,
\begin{equation}
    \label{eq:gam}
    \displaystyle
    \phi_i(t) = t^{l-1} e^{-2 \pi b \text{ERB}(f_i)t} \cos(2 \pi f_i t + c \ln(t)),
\end{equation}
where $l$ and $b$ are Gamma distribution parameters that control the attack and the decay of the impulse response, $c$ is referred to as the chirp parameter which modulates the carrier frequency allowing to slightly modify the instantaneous frequency and $f_i$ is the central frequency of the channel $i$. $\text{ERB}(f_i)$ is a linear transformation of $f_i$ on the Equivalent Rectangular Bandwidth scale \cite{glasberg}:
\begin{equation}
    \label{ERB}
    \displaystyle
    \text{ERB}(f_i) = 24.7 + 0.108 f_i.
\end{equation}

The dictionary $\boldsymbol{D}$ is then created with (\ref{eq:gam2}) in discrete time by striding each filter of length $F_l$ across the sampled signal,
\begin{equation}
    \label{eq:dict}
    \displaystyle
    \boldsymbol{D}^T =
    \begin{bmatrix}
    \boldsymbol{\phi}_0[0] & ... & \boldsymbol{\phi}_0[F_l] & \boldsymbol{0_{1 \times r}} & 0 & ...\\
    \boldsymbol{0_{1 \times r}} & \boldsymbol{\phi}_0[0] & ... & \boldsymbol{\phi}_0[F_l] & \boldsymbol{0_{1 \times r}} & ...\\
    \boldsymbol{0_{1 \times r}} & \boldsymbol{0_{1 \times r}} & \boldsymbol{\phi}_0[0] & ...& \boldsymbol{\phi}_0[F_l] & ...\\
    \vdots & & & & & \vdots \\
    \boldsymbol{\phi}_i[0] & ... & \boldsymbol{\phi}_i[F_l] & \boldsymbol{0_{1 \times r}} & 0 & ...\\
    \boldsymbol{0_{1 \times r}} & \boldsymbol{\phi}_i[0] & ... & \boldsymbol{\phi}_i[F_l] & \boldsymbol{0_{1 \times r}} & ...\\
    \boldsymbol{0_{1 \times r}} & \boldsymbol{0_{1 \times r}} & \boldsymbol{\phi}_i[0] & ...& \boldsymbol{\phi}_i[F_l] & ...\\
    \vdots & & & & & \vdots \\
    \boldsymbol{\phi}_k[0] & ... & \boldsymbol{\phi}_k[F_l] & \boldsymbol{0_{1 \times r}} & 0 & ...\\
    \boldsymbol{0_{1 \times r}} & \boldsymbol{\phi}_k[0] & ... & \boldsymbol{\phi}_k[F_l] & \boldsymbol{0_{1 \times r}} & ...\\
    \boldsymbol{0_{1 \times r}} & \boldsymbol{0_{1 \times r}} & \boldsymbol{\phi}_k[0] & ...& \boldsymbol{\phi}_k[F_l] & ...\\
    \end{bmatrix},
\end{equation}
where $\boldsymbol{0_{1 \times r}}$ is a row vector of zeros of length $r$ (stride size) and $k$ is the number of channels.

\subsubsection{Adaptive LCA (ALCA)}
\label{sssec:grad}

Inspired by the fact that individual channel adaptation occurs in the cochlea through the outer hair cells, we characterized in \cite{alca} each channel $i$ with its own parameters  $l_i$, $b_i$ and $c_i$ instead of sharing the same values among all filters. Hence, (\ref{eq:gam}) becomes:
\begin{equation}
    \label{eq:gam2}
    \displaystyle
    \phi_i(t) = t^{l_i-1} e^{-2 \pi b_i \text{ERB}(f_i)t} \cos(2 \pi f_i t + c_i \ln(t)).
\end{equation}

Moreover, we showed in \cite{alca} the differentiable relationship between the energy function (\ref{eq:energy}) and the modulation parameters of the Gammachirp. 
\begin{align}
    \label{eq:chain}
    \displaystyle
    \begin{aligned}
        \frac{\partial E}{\partial (\boldsymbol{c}|\boldsymbol{b}|\boldsymbol{l})} = & \frac{\partial \frac{1}{2}||\boldsymbol{D}\boldsymbol{a} - \boldsymbol{s}||^2}{\partial \boldsymbol{D}\boldsymbol{a}} \frac{\partial \boldsymbol{D}\boldsymbol{a}}{\partial \boldsymbol{D}} \frac{\partial \boldsymbol{D}}{\partial (\boldsymbol{c}|\boldsymbol{b}|\boldsymbol{l})}\\ & + \lambda \frac{d S(\boldsymbol{a})}{d \boldsymbol{a}} \frac{d \boldsymbol{a}}{d \boldsymbol{v}} \frac{\partial \boldsymbol{v}}{\partial \boldsymbol{D}} \frac{\partial \boldsymbol{D}}{\partial (\boldsymbol{c}|\boldsymbol{b}|\boldsymbol{l})}
    \end{aligned},
\end{align}
where $|$ is the "OR" operator and $\boldsymbol{c}$, $\boldsymbol{b}$, and $\boldsymbol{l}$ are vectors containing the parameters of the filters $c_i$, $b_i$, and $l_i$, respectively. This relationship enabled the implementation of the gradient descent, allowing the gradient flow reaching the membrane potential to backpropagate further towards the Gammachirp's parameters \cite{alca}.

Furthermore, We add in this work a new hyperparameter $\alpha$ that scales the sparsity penalty and decouples the neuron's threshold $\lambda$ in (\ref{eq:thresh}) from the Lagrange multiplier $\lambda$ in (\ref{eq:energy}). The final reported objective function to be minimized by the adaptation of the Gammachirps is:

\begin{equation}
    \label{eq:energyscaled}
    \displaystyle
    E = \frac{1}{2}||\boldsymbol{\hat{s}}-\boldsymbol{s}||^2 + \alpha \lambda S(\boldsymbol{a}).
\end{equation}

\begin{figure}[!t]
\centering
\includegraphics[width=\linewidth]{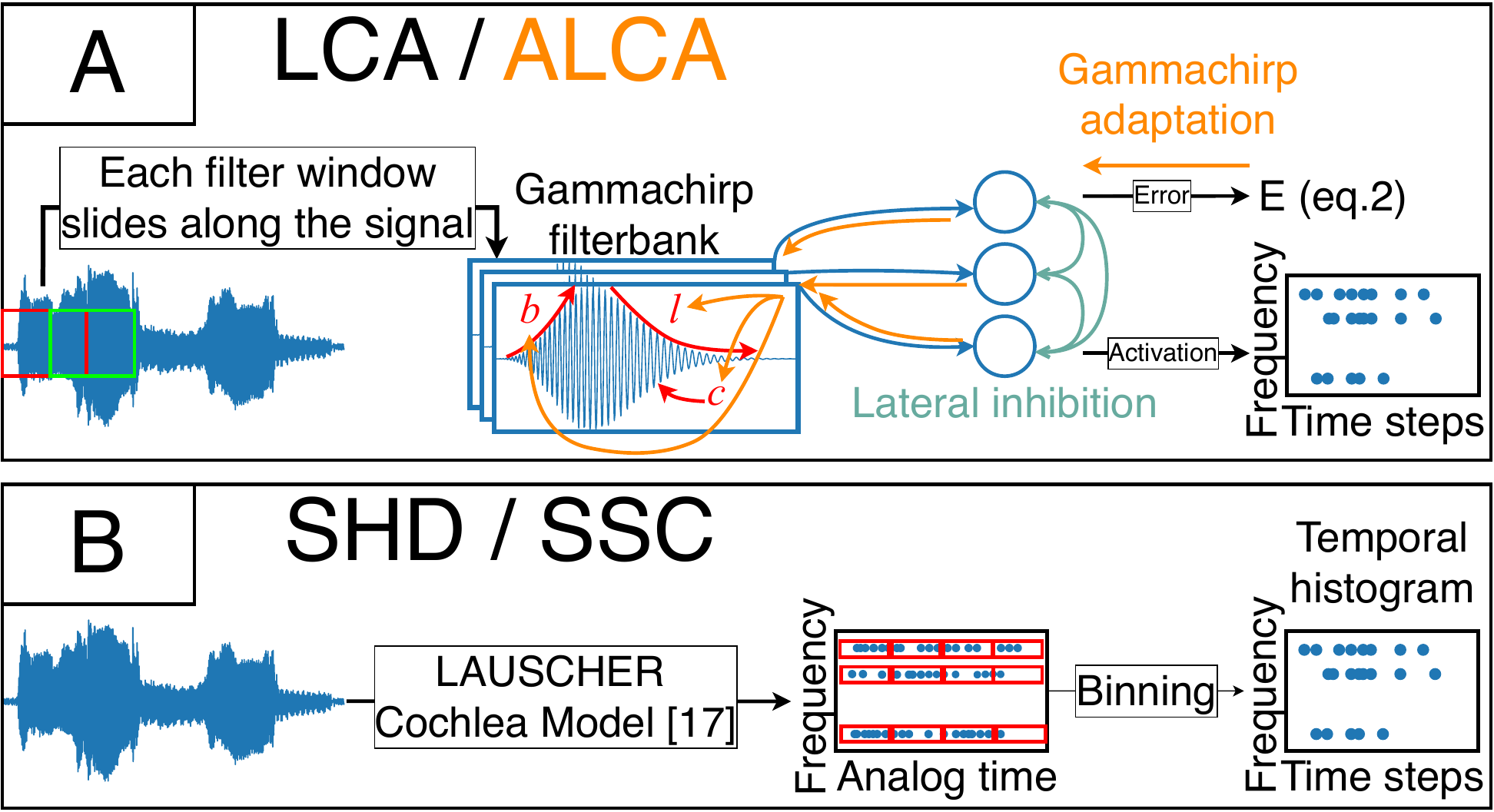}
\caption{ The pipeline to produce LCA/ALCA representations (Block A) and LAUSCHER representations (Block B). Note that the feedback from the energy function $E$ in the block A is only applicable to ALCA.}
\label{fig:lca_sc}
\end{figure}

\begin{table}[!t]
\caption{Hyperparameters of the LCA's Dictionary and Gammatone (GT) as Described in \cite{alca}. HD, and SC Are Described in Sec.~\ref{ssec:audio}.}
\centering
\begin{tabular}{c|c|c|}
\cline{2-3}
\multicolumn{1}{l|}{}                             & \textbf{Parameter} & \textbf{Value} \\ \hline
\multicolumn{1}{|c|}{\multirow{3}{*}{\textbf{Dictionary}}} & $k$         & 700    \\
\multicolumn{1}{|c|}{}                            & $F_l$     & 1024 for HD and 256 for SC  \\
\multicolumn{1}{|c|}{}                            & $r$         & 512 for HD and 128 for SC    \\ \hline
\multicolumn{1}{|c|}{\multirow{3}{*}{\textbf{GT}}}         & $c$         & 0     \\
\multicolumn{1}{|c|}{}                            & $b$         & 1     \\
\multicolumn{1}{|c|}{}                            & $l$         & 4     \\ \hline
\end{tabular}
\label{tab:HP}
\end{table}
The adaptation of Gammachirps in the context of LCA starts from a dictionary composed of Gammatones (GT) atoms, which are a specific type of Gammachirp as defined in \cite{glasberg}, with their parameters indicated in Tab.~\ref{tab:HP}. We then use the truncated backpropagation through time (TBPTT) \cite{TBPTT} and Adamax optimizer \cite{adam} to adapt the modulation parameters. We refer to this algorithm as the Adaptive LCA (ALCA). The LCA and ALCA algorithms are summarized in Block A of Fig.~\ref{fig:lca_sc}

\subsection{Audio Datasets}
\label{ssec:audio}
Two speech classification tasks are studied. First, the Heidelberg digits dataset (HD) \cite{shd} that consists of approximately $10\thinspace{}420$ recordings of spoken digits from zero to nine in both English and German languages including $12$ speakers in total. The sampling rate is $48$ kHz. The dataset is separated into training and test sets. Two speakers are exclusively held for the test set. The remainder of the test set is filled with samples ($5\%$ of the recordings) from speakers also present in the training set. This division allows one to assess a trained network’s ability to generalize across speakers.

The second task is The Google Speech commands dataset (SC), comprising $1$ second long .wav-files with a $16$ kHz sampling rate containing a single English word each \cite{sc}. The dataset contains words spoken by $1\thinspace{}864$ speakers. Specifically, we use the version $2$ of the dataset, consisting of $105\thinspace{}829$ audio files representing 35 classes. For dataset partitioning into training, testing, and validation sets, we employed the same hashing function used in \cite{shd}.

Both datasets are encoded by LAUSCHER and publicly available for benchmarks in \cite{shd}. LAUSCHER representation of HD and SC are called Spiking HD (SHD) and Spiking SC (SSC), respectively. In Block B of Fig.~\ref{fig:lca_sc}, we derive from these representations the temporal histograms of SHD and SSC obtained by binning the analog spike times in time bins of approximately $10$ ms as in \cite{shd} resulting in $128$ time steps.

We also encode both the HD and SC datasets with LCA, selecting $k = 700$ filters to match the number of channels in SHD and SSC. This allows us to compare, in Sec.~\ref{sec:discussion}, our work with state-of-the-art methods that use SHD/SSC with all the channels. We choose the values for the filter length $F_l$ and the stride $r$ that ensure to have similar analysis window (10 ms) as temporal histograms of SHD and SSC. We set the threshold $\lambda = 0.00045$ for HD and $\lambda = 0.0007$ for SC to maintain similar sparsity as SHD and SSC. The values used for the dictionary hyperparameters are in Tab.~\ref{tab:HP}. The representations of HD and SC with this algorithm are called LCA-HD and LCA-SC, respectively. From these LCA representations we derive ALCA representation by adapting the Gammachirps therefore creating two additional representations ALCA-HD and ALCA-SC. All representations are normalized to a $0-1$ range, ensuring all values fall between $0$ and $1$.

In summary we consider 6 representations namely: SHD, SSC, LCA-HD, LCA-SC, ALCA-HD, and ALCA-SC

\subsection{Training Neural Networks}
The data representations produced in this work will serve as the input of neural networks for the speech classification tasks.

\begin{figure}[!t]
\centering
\includegraphics[width=\linewidth]{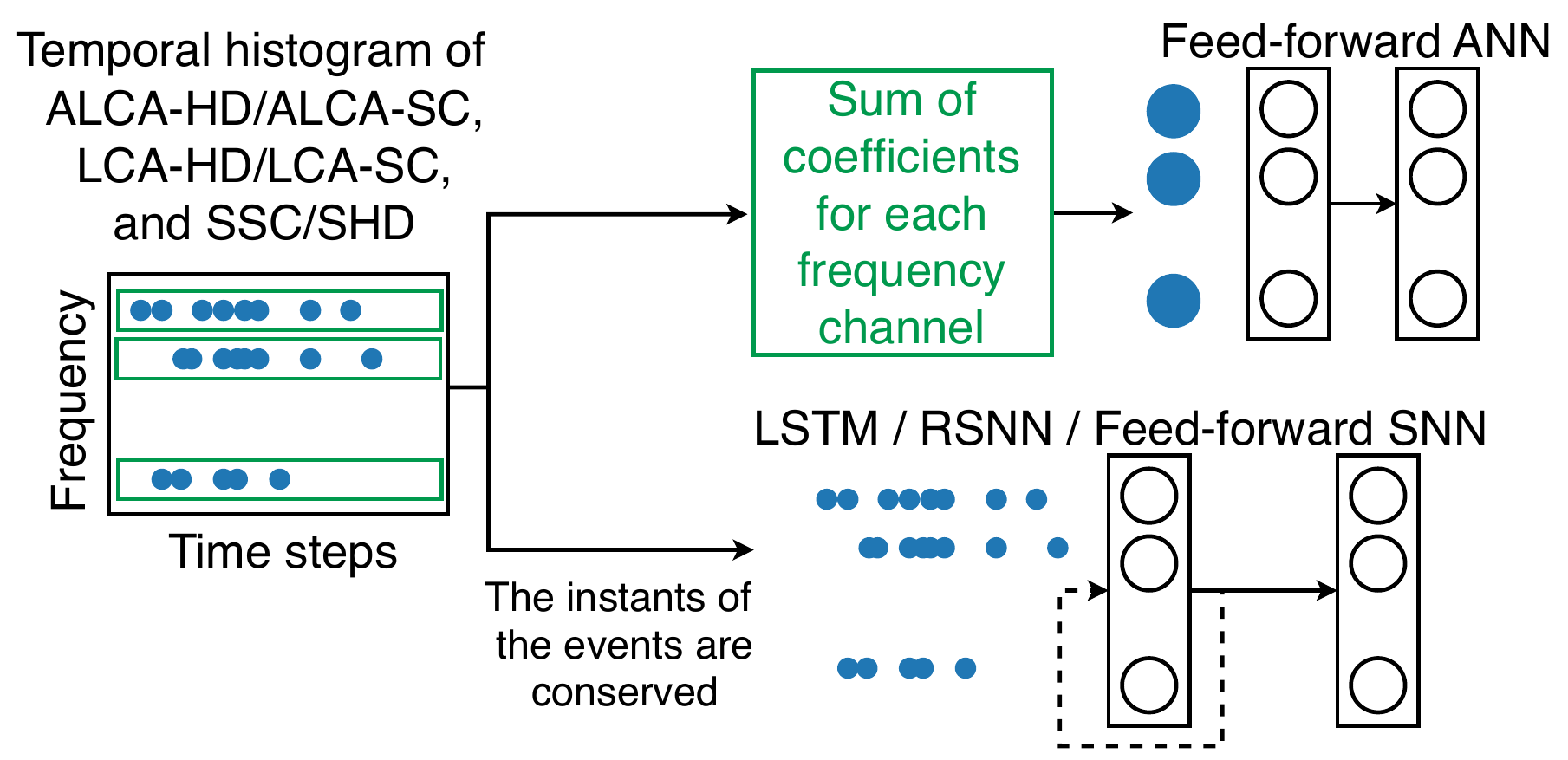}
\caption{Processing of the temporal histogram as input to a feed-forward ANN, LSTM, RSNN and feed-forward SNN. It is to be noted that the sum of coefficients for each frequency channel is used for the feed-forward ANN instead of directly the histogram.}
\label{fig:nets}
\end{figure}

\subsubsection{Artificial Neural Networks (ANNs)}
\label{sec:ann}
To assess the effectiveness of each representation in extracting useful features from speech signals, we use two ANN architectures.

Firstly, we use a feed-forward ANN with fully connected layers, with ReLU activation functions, and with batch normalization \cite{batchnorm} before each layer. We intentionally eliminating the temporal information in the timings of the coefficients from the inputs as illustrated in Fig~\ref{fig:nets}. In fact, we summed the coefficients of each frequency channel over the entire time dimension for each representation. These features will be referred to hereafter as "static features". Therefore, the classification accuracy of models with this architecture reflects the potential of the representations to extract useful features without relying on temporal information.

Secondly, a long short term memory (LSTM) network is used to study the impact of taking into account the temporal information in ALCA and LCA. The inputs to the LSTM consist of temporal histograms of SHD and SSC and the LCA/ALCA representations by keeping the temporal information in the timings of the events (non zero coefficients)(Fig.~\ref{fig:nets}). These features will be called in the following as "temporal features".

Both feed-forward and LSTM architectures use a linear readout as the output layer with the Softmax activation function. In the case of LSTM, the outputs of the readout layer are averaged along all time steps before applying the Softmax function. The models are trained with the Adamax optimizer \cite{adam} and the cross entropy loss using the Pytorch library \cite{pytorch}. For all models, we use the default parameters and initialization of Pytorch unless mentioned otherwise.

\subsubsection{Spiking Neural Networks}
\label{sec:snn}
to compare the energy/power requirements of conventional ANN and SNN, we also implement a fully connected feed-forward SNN and a recurrent SNN (RSNN) both with a current based Leaky Integrate-and-Fire neuron model (Sec.VI in supplemental materials). The last layer of both  architectures is a linear readout. The cross entropy loss is applied on the average of each readout unit over all time steps. The goal of the learning is to minimize the loss function over the entire dataset. To achieve this, a fast sigmoid surrogate gradient descent \cite{superspike} is applied:
\begin{equation}
    \label{eq:fastsigm}
    \displaystyle
    f(x) = \frac{x}{1+\beta|x|},
\end{equation}
where $\beta$ is the steepness parameter. We use backpropagation through time (BPTT) for fully connected networks and truncated BPTT (TBPTT) \cite{TBPTT} for recurrent ones to avoid exploding and vanishing gradients. We use Adamax optimizer with $L_2$ weight regularization.
To prevent early network quiescence, we use a modified form of Glorot initialization \cite{glorot} where the weights are drawn from a normal distribution with custom mean $\mu = multiplier \times \sigma$, as in \cite{SRM}.

\subsection{Hyperparameters Optimization}
\label{sec:hyperopt}
\subsubsection{ALCA}
For ALCA representations, we optimize the learning rate, the mini-batch size, the TBPTT window size, and the scale $\alpha$ by performing an hyperparameter search using the Hyperopt algorithm \cite{hyperopt} with ranges detailed in Tab.~IV in supplemental materials. The search agent minimizes the objective function (\ref{eq:energyscaled}) on the HD and SC test sets after 20 epochs. The search is run for $400$ trials. Each trial is a combination of random values of the hyperparameters. We choose the hyperparameters of the model that produces the most sparse representation.

\subsubsection{Neural Networks}
For neural networks, we optimize the hyperparameters in Tab.~V in supplemental materials using the Hyperopt algorithm. The optimization process involves a search agent running 800 trials for 20 epochs each, aiming to maximize the validation accuracy. Our goal is to identify the model that achieves the highest average validation accuracy while maintaining an acceptable standard deviation below 1\%.

Due to memory limitations, models are initialized with only one seed during the search phase. After completing this initial search, we select the model that demonstrates the best validation accuracy. Practically, we approximate the distribution of validation accuracies for a given model on different seeds to a normal distribution $N(\mu, \sigma)$, where the minimum and maximum values are approximately the $\mu \pm 3 \times \sigma$, resulting in a range of $6 \times \sigma$.

Thus, if the difference between a model's validation accuracy and the best validation accuracy does not exceed $6 \times 1\%$, we consider this model for further evaluation. This approach ensures that if a non-selected model were to achieve the best validation accuracy with a different seed, its standard deviation would necessarily exceed $1\%$. 

Subsequently, we train the selected models using 10 different initialization seeds to further validate their performance. Finally, we choose the model that exhibits the best mean validation accuracy.

\section{EXPERIMENTS AND RESULTS}
\label{sec:rad}
In this section, we present the findings from our comparison of LAUSCHER, LCA, and ALCA across many aspects. 

The first aspect we explore is the relevance of features within each representation. To study the efficacy of feature extraction, we perform a classification using the representations with ANNs. Since all representations are in the time-frequency domain, we initially assess feature importance without relying on time information using fully connected ANNs. Subsequently, we investigate the importance of time information using LSTM networks. This two stage analysis establishes a baseline understanding of how different features contribute to the model’s performance.

Moving to the second aspect of comparison, we examine sparsity and its potential contribution to optimizing neural network computation. For this, we employ SNNs due to the event-based nature of the representations, which aligns well with SNNs' computational principles. Moreover, the inherent sparsity of SNNs, where neurons only compute input currents upon receiving spikes, enables us to demonstrate the impact of sparsity within each representation on neural network computations.

Finally, we include an analysis of power consumption on neuromorphic hardware when solving the classification task using the six representations. This comprehensive study not only highlights the efficiency difference between LAUSCHER and LCA representations but also underscores the impact of the adaptation introduced by ALCA \cite{alca} on both classification and power performance.

\subsection{Static Features}
\label{sec:fe}
Our primary objective is to assess the relevance of feature extraction with LCA and ALCA. To gain insights into the potential for extracting features without relying on temporal information, we first generated a compressed version of the datasets by eliminating all temporal information, as described in Section~\ref{sec:ann}. This means that the sparsity information is also removed. Thus, the sparsity is not subject to evaluation at this stage. Following this, we trained the best fully connected neural network for each representation and dataset as described in Sec.~\ref{sec:ann} and assessed their classification performance.

\begin{figure}[!t]
\centering
\includegraphics[width=\linewidth]{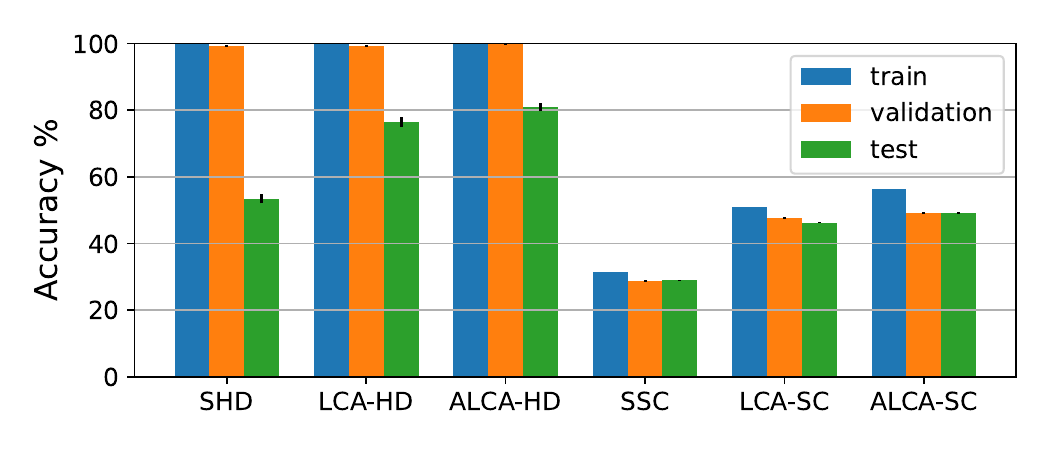}
\caption{Best accuracy achieved by fully connected ANNs with both HD and SC datasets with three representations: LAUSCHER S[HD$|$SC], LCA -[HD$|$SC], and ALCA-[HD$|$SC].}
\label{fig:fc_ann}
\end{figure}

Fig.~\ref{fig:fc_ann} illustrates that while all representations of HD exhibit poor generalization on the test set, ALCA-HD demonstrates superior generalization across speakers compared to LCA-HD, with the latter outperforming SHD. Indeed, ALCA-HD attains the highest test accuracy of $80.87\% \pm 1.22$. Transitioning to the SC dataset, although the overfitting is decreased compared to HD, the overall accuracy experiences a decline. Within this context, ALCA-SC achieves the highest test accuracy of $49.17\% \pm 0.18$, surpassing LCA-SC by almost $3\%$, which leads to approximately $17.27\%$ greater accuracy than SSC.

These results indicate that LCA is a more effective representation than LAUSCHER in the context of static features. Moreover, the ALCA does not compromise classification accuracy. Conversely, ALCA enhances it. The adaptation of Gammachirp's parameters makes the features more discernible and facilitates generalization.
\begin{figure}[!t]
\centering
\subfloat[]{\includegraphics[width=0.5\linewidth]{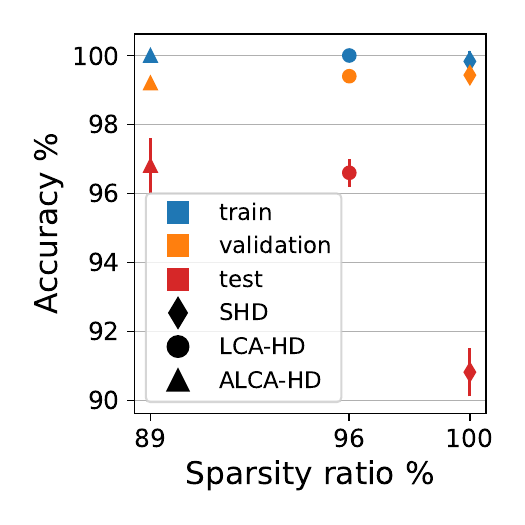}%
\label{fig:lstm_shd}}
\hfil
\subfloat[]{\includegraphics[width=0.5\linewidth]{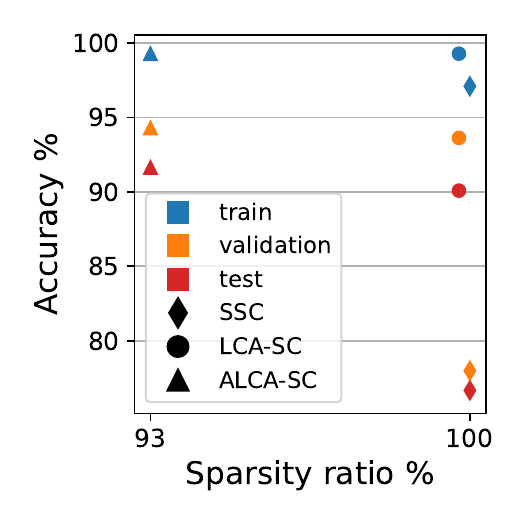}%
\label{fig:lstm_ssc}}
\caption{Best accuracy of LSTM using the three representations –––with temporal information––– with respect to the sparsity ratio (\ref{eq:spratio}) (a) with HD and (b) with SC.}
\label{fig:lstm}
\end{figure}
\begin{figure*}[!t]
\centering
\begin{minipage}[b]{1.0\linewidth}
  \centering
  \centerline{\includegraphics[width=1\linewidth]{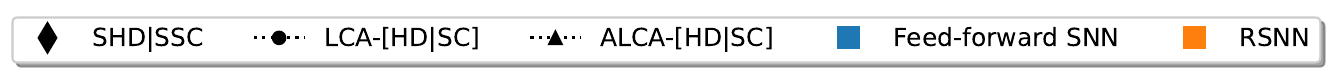}}
\end{minipage}
\hfil
\subfloat[]{\includegraphics[width=0.5\linewidth]{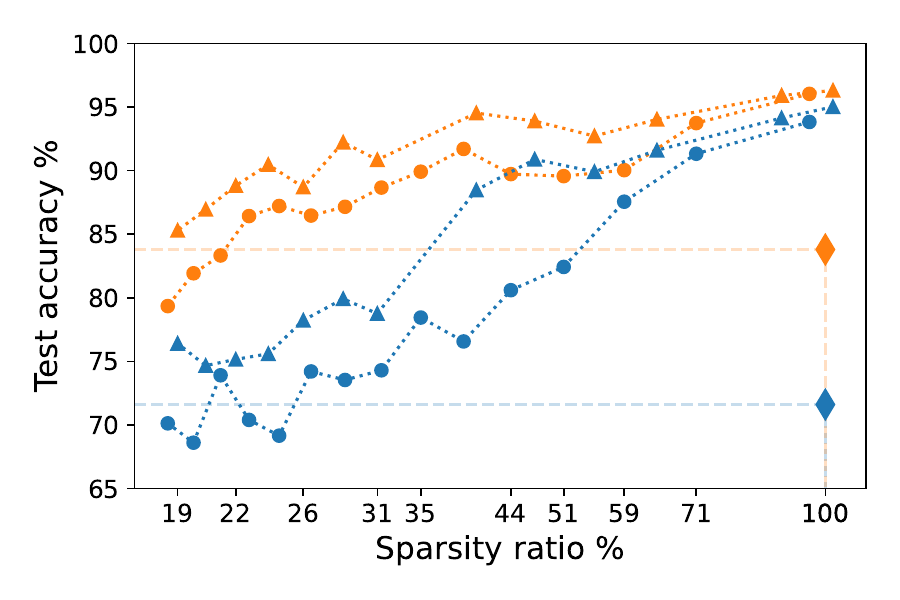}%
\label{fig:snn_shd}}
\subfloat[]{\includegraphics[width=0.5\linewidth]{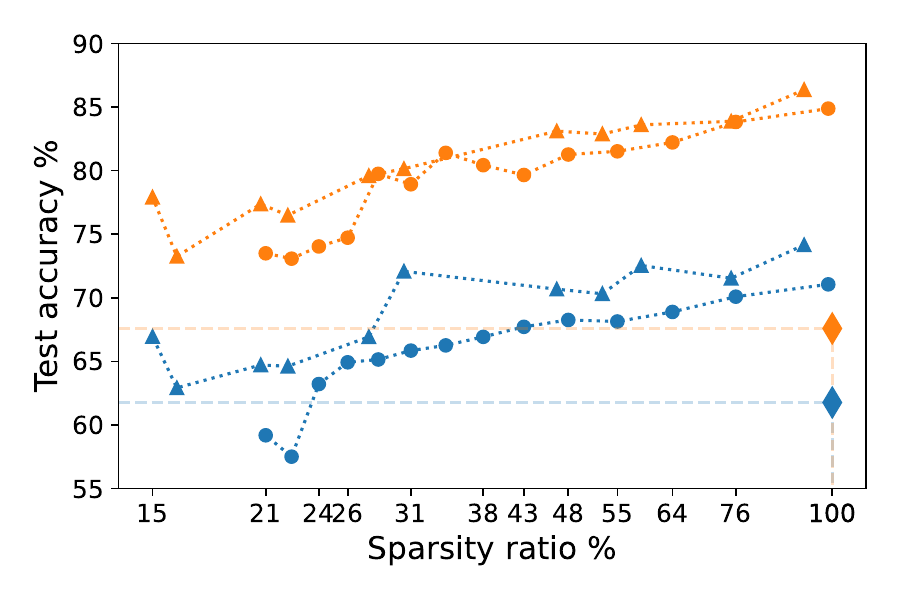}%
\label{fig:snn_ssc}}
\caption{Comparison of the test classification accuracy of Feed-forward and Recurrent SNNs using HD (a) and SC (b). The dotted lines show the trend of the evolution of the test accuracy with respect to the sparsity ratio \% (\ref{eq:spratio}). The chosen values of the hyperparameters of ALCA/LCA and of the SNNs along side the neuron thresholds that give these sparsity ratios are reported in the supplemental materials.}
\label{fig:snn}
\end{figure*}
\begin{figure}[!t]
\centering
\begin{minipage}[b]{1.0\linewidth}
  \centering
  \centerline{\includegraphics[width=1\linewidth]{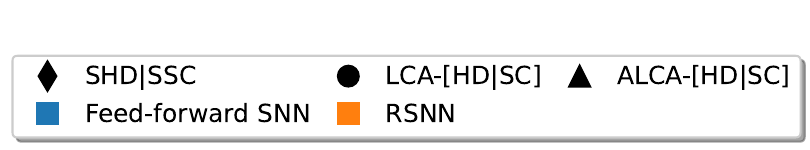}}
\end{minipage}
\hfil
\subfloat[]{\includegraphics[width=0.5\linewidth]{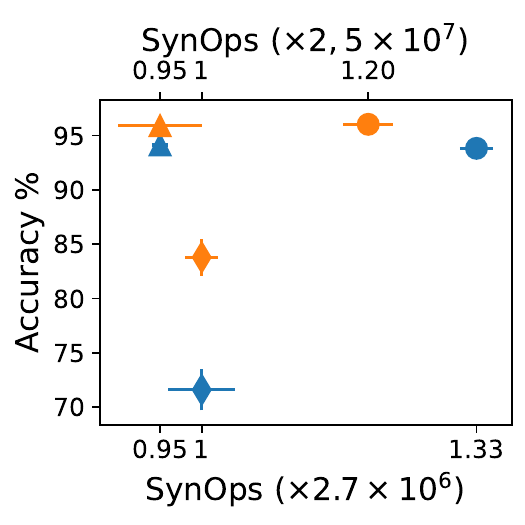}%
\label{fig:synops_shd}}
\subfloat[]{\includegraphics[width=0.5\linewidth]{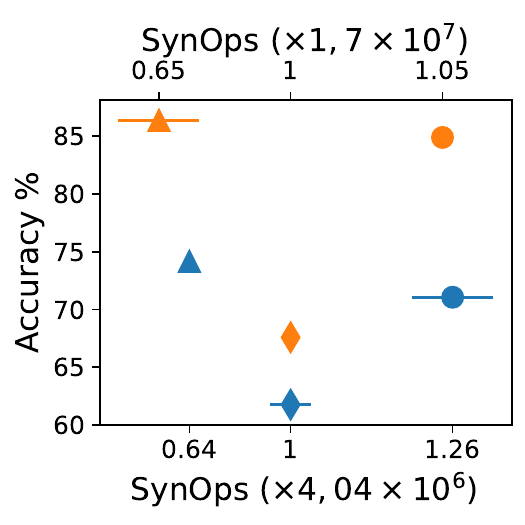}%
\label{fig:synops_ssc}}
\caption{Test classification accuracy of SNNs using HD (a) and SC (b) versus the number of synaptic operations (SynOps). In both figures, the top x-axis for RSNNs is scaled by the SynOps of the RSNN with LAUSCHER representation of the dataset. The bottom x-axis for feed-forward SNNs is scaled by the SynOps of the feed-forward SNN with LAUSCHER representation of the dataset.}
\label{fig:synops}
\end{figure}

\subsection{Importance of Temporal Features}
\label{sec:ti}
It was observed in \cite{shd} that the temporal information was necessary to reduce the overfitting on the SHD and SSC representations. To assess whether this situation is the same for LCA and ALCA, we trained LSTMs using the coefficients obtained from LCA and ALCA as well as temporal histograms derived from the spike activity of SHD and SSC as described in Sec.~\ref{sec:ann}. This temporal processing will keep the sparsity of representations and will give an insight on the balance between the classification performance and the sparsity of each representation. The hyperparameter optimization was conducted for each representation and dataset. Refer to the supplemental materials for the chosen values of the hyperparameters of the ALCA and of the LSTMs.

In Fig.~\ref{fig:lstm} we show the optimal classification performance for each setup with respect to the sparsity ratio $r$:
\begin{equation}
    \label{eq:spratio}
    \displaystyle
    r = \frac{\#_\text{c}([\text{A}]\text{LCA}-(\text{HD}|\text{SC}))}{\#_\text{s}(\text{S}(\text{HD}|\text{SC}))},
\end{equation}
where $\#_\text{c}(.)$ is the average number of non-zero coefficients in the considered representation (ALCA or LCA) computed from the test set of the considered dataset, and $\#_\text{s}(.)$ is the average spike count of the LAUSCHER representation computed from the test set of the considered dataset.

In fig.~\ref{fig:lstm}(a), with the comparable sparsity as SHD, LCA-HD leads to a test accuracy of $96.60\% \pm 0.4$, representing a $6\%$ improvement over SHD. Furthermore, the adaptation of Gammachirps filter bank in ALCA-HD achieves a $11\%$ sparsity increase compared to SHD without compromising the accurate classification achieved by LCA-HD. In fig.~\ref{fig:lstm}(b), LCA-SC achieves, with approximately the same sparsity as SSC, a test accuracy of $90.08\% \pm 0.3$, showing a substantial $13\%$ improvement over SSC. ALCA-SC is $7\%$ sparser than LCA-SC and SSC and it leads to the best test accuracy of $91.63\% \pm 0.1$. These results indicate that LCA exhibits more accurate classification compared to LAUSCHER, suggesting its efficacy in processing speech information. Furthermore, ALCA not only outperforms LAUSCHER in terms of speech classification accuracy but also emerges as the sparsest representation, highlighting its efficiency in encoding speech signals. 

The temporal information proves beneficial as all representations in fig.~\ref{fig:lstm} exhibit enhanced generalization compared to their non-temporal counterparts in fig.~\ref{fig:fc_ann}

\subsection{How Sparse Are LAUSCHER, LCA, and ALCA?}
\label{sec:sparsity}
Having demonstrated the relevance of LCA and ALCA as effective feature extractors in the context of speech classification, along with the significance of incorporating temporal information, we sought to evaluate the performance of Spiking Neural Networks (SNNs). Our focus is to assess how well SNNs perform when applied to these established representation methods, particularly in relation to the sparsity of these representations. We systematically varied the threshold levels of neurons within LCA, resulting in a spectrum of average non-zero coefficients spanning approximately from $20\%$ to $100\%$ of the average spike count observed in the SHD and SSC datasets in the previous section.

For each sparsity ratio (\ref{eq:spratio}), we optimize the hyperparameters in Tab.~IV in supplemental materials to adapt the Gammachirp filter bank, and the hyperparameters in Tab.V in supplemental materials to train SNNs. For more details refer to Sec.~\ref{sec:hyperopt}. The chosen values are in Sec.VII of supplemental materials. Fig.~\ref{fig:snn} presents a comparison of the best test classification accuracy achieved for each sparsity ratio. Fig.~\ref{fig:snn}(a) illustrates that, with feed-forward SNNs, the best test accuracy using SHD representation is $71.61\% \pm 1.9$. The LCA-HD, with a similar sparsity, yields significantly improved results, achieving a higher best test accuracy of $93.83\% \pm 0.6$. Notably, the ALCA-HD representation, with a comparable sparsity to SHD, demonstrates even better performance, attaining the highest best test accuracy of $95.02\% \pm 0.7$. Furthermore, feed-forward SNNs with LCA-HD achieve a comparable best test accuracy to that achieved with SHD, but with only approximately $26\%$ of SHD's sparsity, making LCA-HD nearly $4$ times sparser than SHD. On the other hand, feed-forward SNNs with ALCA-HD achieve even better test accuracy with just $19\%$ of SHD's sparsity, resulting in ALCA-HD being more than $5$ times sparser than SHD. 
Although the disparity between ALCA-HD and LCA-HD decreases at high sparsity ratios, feed-forward SNNs with ALCA-HD achieve better test accuracy than those with LCA-HD across all sparsity ratios. Consistent results are observed with RSNNs where the top test accuracies with SHD, LCA-SHD, and ALCA-SHD representations are respectively of $83.80 \pm 1.7$, $96.04 \pm 0.9$, and $96.31 \pm 0.8$. Notably, for comparable test accuracy, LCA-HD and ALCA-HD demonstrate approximately $5$ times greater sparsity than SHD.

All these findings are supported by experiments on SSC, as shown in Fig.~\ref{fig:snn}(b), where the differences between SSC, LCA-SC, and ALCA-SC is more pronounced around the $100\%$ sparsity ratio. Notably, RSNNs with LCA-SC achieve better test accuracy than SSC even at just $20\%$ of SSC's sparsity. ALCA-SC pushes this even further, reducing the sparsity to just $15\%$ of SSC's sparsity, where RSNNs continue to outperform SSC in test accuracy.

These collective results underscore that LCA is a more effective coding method than LAUSCHER, and with adapted Gammachirps, ALCA further improves upon this efficacy.

\subsection{Spiking Activity}
\label{sec:spact}
Another aspect of speech representations that we focus on is their impact on the activity of the processing SNN. This focus arises from the fact that improving the sparsity of a representation does not necessarily mean reducing the activity of the processing neural network. Thus, we compare the activity of SNNs using LAUSCHER and LCA representations, which exhibit the same representation sparsity (sparsity ratio = $100\%$ in Fig.~\ref{fig:snn}). To further assess the influence of the gained sparsity with ALCA, we also compare SNNs using ALCA representations derived from the LCA representations.

The classification accuracy of SNNs using these representations is evaluated with respect to the number of synaptic operations (SynOps), as illustrated in Fig.~\ref{fig:synops}. For both datasets, feed-forward SNNs utilizing LCA achieve better classification accuracy than feed-forward SNNs using the reference SHD and SSC representations. However, we observe that the LCA representations generated more activity in the SNN despite having a similar sparsity as LAUSCHER representations. More precisely, feed-forward SNNs with LCA-HD produce $33\%$ more SynOps than with SHD in Fig.~\ref{fig:synops}(a), and SynOps using LCA-SC are $26\%$ higher than those using SSC in Fig.~\ref{fig:synops}(b). Moreover, RSNNs confirm these results where LCA-HD causes an increase of $20\%$ in SynOps compared to SHD in Fig.~\ref{fig:synops}(a). Similarly, the RSNN with LCA-SC produces $5\%$ more SynOps compared to RSNN with SSC in Fig.~\ref{fig:synops}(b). These results highlight differences in network activity even when LCA and LAUSCHER representations have a similar sparsity.

On the other hand, Fig.~\ref{fig:synops} demonstrates that ALCA is the most efficient representation among the three. In fact, ALCA-HD combines the accuracy achieved with LCA-HD and the sparsity of the spiking activity achieved with SHD Fig.~\ref{fig:synops}.a. Furthermore, ALCA-SC exhibits a reduction of $25\%$ in SynOps compared to SSC in Fig.~\ref{fig:synops}(b) with both feed-forward SNNs and RSNNs. This sparsity in SNNs activity improvement with ALCA do not compromise the test accuracy. In both cases ALCA leads to the most accurate classification with the sparsest SNN's activity.

The study confirms that LCA consistently outperforms LAUSCHER regarding test accuracy across all evaluated cases, albeit with a higher SynOps cost. However, ALCA 
addresses this trade-off by enhancing the activity efficiency of the processing SNN, without compromising the classification accuracy.

         
\begin{table*}[!t]
\centering
\caption{Comparison of Dynamic Power/energy consumption per inference and inference time of a RSNN of 1024 Neurons alongside the Best Test Accuracy Achieved for HD and SC Datasets across the Three Established Representations.}
\begin{tabular}{|r|ll|ll|ll|ll|ll|ll|}
\hline
\textbf{Dataset} & \multicolumn{2}{c|}{\textbf{SHD}} & \multicolumn{2}{c|}{\textbf{LCA-HD}} & \multicolumn{2}{c|}{\textbf{ALCA-HD}} & \multicolumn{2}{c|}{\textbf{SSC}} & \multicolumn{2}{c|}{\textbf{LCA-SC}} & \multicolumn{2}{c|}{\textbf{ALCA-SC}} \\ \hline
\textbf{Hardware} & \multicolumn{1}{l|}{\textbf{GPU}} & \textbf{Loihi 2} & \multicolumn{1}{l|}{\textbf{GPU}} & \textbf{Loihi 2} & \multicolumn{1}{l|}{\textbf{GPU}} & \textbf{Loihi 2} & \multicolumn{1}{l|}{\textbf{GPU}} & \textbf{Loihi 2} & \multicolumn{1}{l|}{\textbf{GPU}} & \textbf{Loihi 2} & \multicolumn{1}{l|}{\textbf{GPU}} & \textbf{Loihi 2} \\ \hline
\textbf{Dynamic Power (W)} & \multicolumn{1}{l|}{$54$} & $0.015$ & \multicolumn{1}{l|}{$54$} & $0.021$ & \multicolumn{1}{l|}{$54$} & $\boldsymbol{0.013}$ & \multicolumn{1}{l|}{$52$} & $0.006$ & \multicolumn{1}{l|}{$52$} & $0.028$ & \multicolumn{1}{l|}{$52$} & $\boldsymbol{0.004}$ \\ \hline
\textbf{Inference Time (s)} & \multicolumn{1}{l|}{$0.20$} & $5.42$ & \multicolumn{1}{l|}{$0.20$} & $5.45$ & \multicolumn{1}{l|}{$0.20$} & $\boldsymbol{5.12}$ & \multicolumn{1}{l|}{$0.20$} & $4.43$ & \multicolumn{1}{l|}{0.20} & $5.94$ & \multicolumn{1}{l|}{$0.20$} & $\boldsymbol{4.31}$ \\ \hline
\textbf{Dynamic Energy (J)} & \multicolumn{1}{l|}{$10.8$} & $0.081$ & \multicolumn{1}{l|}{$10.8$} & $0.114$ & \multicolumn{1}{l|}{$10.8$} & $\boldsymbol{0.067}$ & \multicolumn{1}{l|}{$10.4$} & $0.027$ & \multicolumn{1}{l|}{$10.4$} & $0.166$ & \multicolumn{1}{l|}{$10.4$} & $\boldsymbol{0.017}$ \\ \hline
\multirow{2}{*}{\textbf{\begin{tabular}[c]{@{}r@{}}Quantized Model \\ Test Accuracy (\%)\end{tabular}}} & \multicolumn{2}{c|}{\multirow{2}{*}{$83.77 \pm 1.2$}} & \multicolumn{2}{c|}{\multirow{2}{*}{$\boldsymbol{94.63\pm 0.8}$}} & \multicolumn{2}{c|}{\multirow{2}{*}{$94.38 \pm 1.0$}} & \multicolumn{2}{c|}{\multirow{2}{*}{$68.53 \pm 0.4$}} & \multicolumn{2}{c|}{\multirow{2}{*}{$84.44 \pm 0.5$}} & \multicolumn{2}{c|}{\multirow{2}{*}{$\boldsymbol{86.40 \pm 0.4}$}} \\
 & \multicolumn{2}{c|}{} & \multicolumn{2}{c|}{} & \multicolumn{2}{c|}{} & \multicolumn{2}{c|}{} & \multicolumn{2}{c|}{} & \multicolumn{2}{c|}{} \\ \hline
\end{tabular}
\label{tab:power}
\end{table*}
\subsection{Power/Energy Efficiency on Loihi 2}
\label{sec:powereff}
We conduct a power/energy consumption benchmark on Nvidia TITAN XP GPU and on Loihi 2. The experiments replicate those in Fig.\ref{fig:synops}, incorporating the quantization constraints specific to Loihi 2. We train the models with the Lava Deep Learning library (\texttt{Lava-DL SLAYER}) \cite{lava-dl}. This library leverages a PyTorch implementation as a backend while considering Loihi 2 hardware constraints.

The inference on the GPU utilizes (\texttt{Lava-DL SLAYER}), and its efficiency is measured using \texttt{nvidia-smi} on Linux with kernel version \texttt{5.4.0-7634}, \texttt{Python 3.8.1} and \texttt{PyTorch 2.0.0}. Loihi’s efficiency is measured using the power probes of \texttt{Lava 0.9.0} on \texttt{Oheogultch} board \texttt{ncl-ext-og-05}.

The dynamic power/energy refers to the power/energy consumed by the hardware while the system is actively running the workload. Given our primary objective of enhancing speech representation sparsity and the resulting activity sparsity in the SNN, we focus on the dynamic power/energy per inference. Studying the static power/energy is beyond the scope of this paper, as it is related to the hardware itself and not to the neural network running on it.

The dynamic power consumption results in Tab.~\ref{tab:power} reveal the remarkable efficiency of Loihi 2 compared to the GPU. While the model on the GPU consumes $54$ W, the Loihi implementation demonstrates significantly lower power consumption for all the cases. 

ALCA-HD enables the recurrent neural network to achieve $10\%$ higher classification accuracy compared to using SHD, but this improvement comes at the cost of a $40\%$ increase in power consumption. Nevertheless, ALCA-HD provides a good accuracy-power trade-off, offering high classification accuracy comparable to that of LCA-HD while maintaining low power consumption levels similar to SHD. Moreover, ALCA-SC outperforms both SSC and LCA-SC in terms of classification accuracy and is also the representation that results in the lowest dynamic power consumption for the recurrent neural network.

Note that the inference time is more than $10$ times slower on Loihi 2 than on GPU. This is due to the fast input/output communication that has not been yet activated in the version of Lava library we use in this experiment. Despite the slower inference on Loihi 2, the dynamic energy consumption is still lower on Loihi 2 than on GPU by 2 to 3 orders of magnitude.

These findings highlight the potential of using ALCA for efficient and low-power speech classification tasks when implemented on Intel's Loihi 2.
\begin{table*}[!t]
    \caption{Results of Prior Speech Classification Benchmarks Using LAUSCHER Representation with SNNs Compared to this Work. Note that for Each Work We Report the Highest Achieved Test Accuracy.}
    \centering
    \begin{tabular}{|c|c|c|c|}
        \hline
         \textbf{Sparse Representation}& \textbf{Network} & \textbf{HD Accuracy} \%& \textbf{SC Accuracy} \%\\
         \hline
         LAUSCHER \cite{snn_delay1}&Fully connected SNN with learned delays  & $95.07 \pm 0.24$ & $80.69 \pm 0.21$\\
         \hline
         \multirow{2}{*} {LAUSCHER \cite{rsnn_adapt}}&RSNN with an adaptation mechanism between & \multirow{2}{*}{$94.6$} & \multirow{2}{*}{$77.40$}\\
         &subthreshold and threshold regimes & & \\
         \hline
         LAUSCHER \cite{snn_delay2}&Feed-forward SNN with adaptive axonal delays  & $92.45$ & - \\
         \hline
         \multirow{2}{*}{LAUSCHER \cite{snn_attention}}& Feed-forward SNN with spatio-temporal filters & \multirow{2}{*}{$92.36$} & \multirow{2}{*}{-} \\
         & and attention & & \\
         \hline
         LAUSCHER \cite{rsnn_attention}&RSNN with temporal attention & $91.08$ & - \\
         \hline
         LAUSCHER \cite{snn_delay3}&Feed-forward SNN with random dendritic delays & $90.88$ & - \\
         \hline
         LAUSCHER (this work)& RSNN  & $83.80\% \pm 1.7$ & $67.59 \pm 0.3$ \\
         \hline
         LCA (this work)& RSNN & $96.04\% \pm 0.9$ & $84.88 \pm 0.3$ \\
         \hline
         ALCA (this work)& RSNN & \boldmath $96.31\% \pm 0.5$ & \boldmath $86.36 \pm 0.5$ \\
         \hline
    \end{tabular}
    \label{tab:shd_review}
\end{table*}

\section{DISCUSSION}
\label{sec:discussion}
This study demonstrates the applicability of LCA and ALCA as sparse representations for speech classification. 

We emphasize the ALCA's superior ability to generalize on the test set compared to LCA and LAUSCHER in speech classification. The LCA effectively captures discriminating and informative features by filtering out irrelevant ones and focusing on the most significant ones, thereby improving speech classification compared to LAUSCHER. Additionally, the LCA demonstrates superior generalization, attributed to its reduced sensitivity to extraneous information. This enhanced robustness enables it to handle variations and changes in input data more effectively than LAUSCHER. 

Building upon the advantages of the LCA, the ALCA further enhances these qualities. The improved accuracy and the generalization ability of the ALCA demonstrates that the enhanced reconstruction quality showed in \cite{alca} is not due to reconstructing the noise in the data but is a result of finding the Gammachirp filters that better represent the data so that with fewer non-zero coefficients, the speech classification is easier. In fact, the quality metric used in \cite{alca} is the mean squared error (MSE) between the original signal and the reconstructed signal. While ALCA achieved a lower MSE compared to LCA, this does not necessarily indicate an improvement in quality. A model that reproduces the noise in the original signal could achieve a lower MSE than one that does not replicate the noise. However, this study demonstrated that ALCA does not fall into this category. Notably, speech signals represented by ALCA are more easily discriminated than those represented by LCA.

In addition to their utility for speech classification, both LCA and ALCA prove beneficial for temporal tasks such as the HD and SC classification. In fact, LSTMs (Fig.\ref{fig:lstm}) and recurrent and feed-forward SNNs (Fig.\ref{fig:snn}) achieve better classification accuracy using temporal information than non-temporal feed-forward ANNs (Fig.\ref{fig:fc_ann}). This suggests that recurrent synapses and the dynamics of spiking neurons optimize accuracy and generalization by using the temporal interval between non-zero coefficients.

Furthermore, the research underscores that enhancing the sparsity of a data representation does not inherently imply a reduction of the processing neural network's activity. LCA is an example, as its sparsity alone does not result in less or comparable activity than LAUSCHER in the processing neural network. However, ALCA mitigates this issue, improving the sparsity of the processing SNN across various architectures and datasets, thus reducing power consumption on Loihi 2.

Finally, ALCA stands out compared to other considered approaches enhancing the classification accuracy of LAUSCHER representations (SHD and SSC), which often require an increased model complexity. In fact, previous works have improved the classification accuracy by adding learnable delays, differential equations to model an adaptation variable between threshold and sub-threshold voltage, adaptive axonal delays, random dendritic delays, temporal attention, and spatio-temporal filters and attention (see Tab.~\ref{tab:shd_review}). However, none of these studies achieved the test accuracy demonstrated by ALCA, which we tested with a basic current based LIF neuron model. ALCA's effectiveness in achieving high accuracy and low power consumption with a simplified model underscores its efficiency and superiority in this study's context.

\section{CONCLUSION}
\label{sec:conclusion}
In this study, we evaluated ALCA \cite{alca} and LCA as sparse representations for speech signals. Our evaluation involved a thorough comparison with a reference representation produced by the LAUSCHER cochlea model \cite{shd} in the context of neuromorphic speech classification tasks.

The obtained results highlight the superior effectiveness of ALCA over LCA and, in turn, the superiority of LCA over the LAUSCHER representation. In fact, LCA showcased high generalization capabilities in speech classification, emphasizing its ability to capture more discriminating and informative features than LAUSCHER. Building upon these strengths, ALCA improved the classification accuracy and the generalization by optimizing Gammachirp filters for better data representation. Moreover, the results consistently indicated that either LCA leads to a more accurate classification than LAUSCHER for the same representation sparsity or LCA is a sparser representation than LAUSCHER representation for comparable classification accuracy.

However, a critical observation emerged from the study: having two different sparse representations with comparable sparsity does not guarantee comparable activity in the processing neural network. Specifically, LCA produced more activity in the processing network than LAUSCHER representation, even though both representations had a comparable sparsity. This observation underscores the importance of analyzing the sparsity of the processing network rather than solely focusing on the representation's sparsity.

ALCA effectively addressed this issue by enhancing the sparsity of the processing network across various architectures and datasets, thereby reducing power consumption on Loihi 2. Significantly, the improvements achieved by ALCA in terms of the reconstruction quality and the sparsity, as documented in \cite{alca}, did not compromise classification performance. On the contrary, ALCA achieved high accuracy compared to the state-of-the-art and low power consumption with a simplified spiking neuron model. This finding underscores the efficiency and superiority of ALCA in speech classification, positioning it as a promising candidate for spike-based computations in audio-related tasks.

In summary, ALCA not only emerges as an alternative representation for tasks requiring high reconstruction quality \cite{alca} but also establishes itself as an efficient solution for low-power speech classification tasks. Its dual potential makes ALCA a promising solution for spike-based computations in various speech-related applications.

\section*{ACKNOWLEDGMENTS}
\label{sec:ack}
The authors would like to thank the "Fonds de recherche du Québec - Nature et technologies" and "Natural Sciences and Engineering Research Council of Canada" for funding this research. We extend our appreciation to NVIDIA for donating the GTX1080 and Titan Xp GPUs. We would also like to thank Digital Research Alliance of Canada for providing access to servers equipped with the necessary CPUs and GPUs resources, enabling us to conduct our experiments effectively. Finally, we thank Intel for giving us access to Loihi 2 and Andreas Wild for his insights on the power/energy benchmark.

\bibliographystyle{IEEEtran}
\bibliography{refs}
\end{document}